\documentclass[twocolumn,aps,superscriptaddress]{revtex4-1}
\usepackage{amsmath}
\usepackage{bm}
\usepackage{graphicx}
\usepackage{mathrsfs}
\usepackage{multirow}
\usepackage[normalem]{ulem}
\usepackage{graphicx}
\usepackage{booktabs, ctable}
\usepackage{xcolor, framed}
\usepackage[colorlinks=true,
            linkcolor=red,
            citecolor=blue,
            urlcolor=black]{hyperref}
            
\graphicspath{{./figs/}}

\begin{document}

\title{From hyperon--nucleon interactions to deuteron--hyperon femtoscopy}

 \author{Jiaxing Zhao}
 \email{jzhao@itp.uni-frankfurt.de}
 \affiliation{Helmholtz Research Academy Hesse for FAIR (HFHF), GSI Helmholtz Center for Heavy Ion Physics, Campus Frankfurt, 60438 Frankfurt, Germany}
 \affiliation{Institut f\"ur Theoretische Physik, Johann Wolfgang Goethe-Universit\"at,Max-von-Laue-Straße 1, D-60438 Frankfurt am Main, Germany}
\date{\today}

\begin{abstract}
I investigate the low-energy scattering and femtoscopic correlation functions of the $d-\Lambda$, $d-\Sigma$, and $d-\Xi$ systems within a microscopic folding approach. The effective deuteron--hyperon interactions are constructed by folding the HAL-QCD hyperon--nucleon potentials with the deuteron wave function, while the spin and isospin structures are treated through Wigner-$6j$ recoupling coefficients. Using the resulting interactions, I calculate the scattering parameters and momentum correlation functions for all spin channels. No bound states are found for the $d-\Lambda$, $d-\Sigma$, or $d-\Xi$ systems. Nevertheless, the $d-\Lambda$ correlation exhibits a pronounced low-momentum enhancement associated with a large scattering length and
a near-threshold pole, whereas the $d-\Sigma$ correlation is suppressed by its predominantly repulsive interaction. The neutral $d-\Xi^{0}$ system shows only a moderate enhancement, while the charged $d-\Xi^{-}$ correlation is strongly amplified by the attractive Coulomb interaction. I further investigate feed-down effects from $\Sigma^{0}$, $\Sigma(1385)$, and $\Xi$ decays using Monte Carlo response matrices and demonstrate that these decays clearly modify the observable $d-\Lambda$ correlation. The results provide quantitative predictions for future femtoscopic measurements and establish deuteron--hyperon correlations as a sensitive probe of hyperon--nucleus interactions.
\end{abstract}

\maketitle

\section{Introduction}
The interaction between hyperons and nucleons plays a fundamental role in nuclear and hadronic physics. It governs the formation and structure of hypernuclei, determines the properties of strange nuclear matter, and provides essential input for understanding the composition and equation of state of neutron stars~\cite{Lattimer:2015nhk,Schaffner-Bielich:2008zws}. While nucleon--nucleon interactions have been determined with high precision through abundant scattering data, direct experimental information on the hyperon--nucleon ($Y-N$, with $Y=\Lambda,\Sigma,\Xi$) interaction remains rather limited because of the short lifetime of hyperons~\cite{Sechi-Zorn:1968mao,Alexander:1968acu}. Consequently, considerable efforts have been devoted to constraining the $Y-N$ interaction using phenomenological meson-exchange models~\cite{Holzenkamp:1989tq,Stoks:1999bz,Rijken:2010zzb}, chiral effective field theory~\cite{Haidenbauer:2013oca,Haidenbauer:2019boi}, and lattice QCD~\cite{Nemura:2017vjc,HALQCD:2019wsz}.

Since hypernuclei are formed by binding one or more hyperons to ordinary nuclei, their binding energies, excitation spectra, and decay properties are determined by the underlying hyperon--nucleus interaction.
Therefore, hypernuclei provide a unique laboratory for probing the $Y-N$ interaction, especially its spin-dependent components~\cite{Gal:2016boi,Botta:2016kqd}. 
The first step beyond the elementary $Y-N$ system is the interaction between a hyperon and the lightest bound nucleus, the deuteron. As the simplest hyperon--nucleus system, the deuteron--hyperon ($d-Y$) interaction provides a direct connection between the underlying $Y-N$ force and the structure of hypernuclei~\cite{Haidenbauer:2020uew}. 
The hypertriton, $\rm ^3_\Lambda H$, has long been discussed as a weakly bound $d-\Lambda$ cluster, although its microscopic structure is more accurately described within a three-body $\Lambda NN$ framework. Its extremely small $\Lambda$ separation energy nevertheless makes it highly sensitive to the low-energy $d-\Lambda$ interaction.
Similarly, the existence and properties of light $\Sigma$ and $\Xi$ hypernuclei are governed by the corresponding $d-\Sigma$ and $d-\Xi$ interactions. Precise knowledge of deuteron--hyperon interactions is therefore indispensable for developing a quantitative description of light hypernuclei.

Femtoscopic correlation has emerged as a powerful experimental tool for probing hadron--hadron interactions~\cite{Lisa:2005dd,Ohnishi:2016elb,Fabbietti:2020bfg,ALICE:2020mfd,Liu:2024uxn,Zhao:2026tpj}. The correlation function at small relative momentum is determined jointly by the emission source and the final-state interaction, enabling the extraction of scattering parameters even when direct scattering experiments are impractical~\cite{Lednicky:2005tb,Lisa:2005dd}. This technique has been successfully applied to nucleon--nucleon~\cite{ALICE:2018ysd}, hyperon--nucleon~\cite{STAR:2005rpl,ALICE:2018ysd,ALICE:2019buq,ALICE:2019hdt,STAR:2018uho}, and hyperon--hyperon~\cite{STAR:2014dcy,ALICE:2018ysd,ALICE:2022uso} systems in high-energy collisions. Extending femtoscopy to deuteron--hyperon pairs offers a complementary approach for investigating effective hyperon--nucleus interactions. Since the deuteron serves as the lightest composite nucleus, the measured $d-Y$ correlation functions are directly sensitive to the low-energy $d-Y$ scattering amplitude and can provide independent constraints on the existence of shallow bound or virtual states, as well as on the corresponding scattering length and effective range.

Experimentally, deuterons and strange baryons are abundantly produced in heavy-ion collisions over a broad energy range. $d-\Lambda$ correlation has been measured in Au+Au collisions at $\sqrt{s_{\rm NN}}=3~\rm GeV$ by the STAR experiment at RHIC~\cite{STAR:2025jwe}.
It will be also particularly feasible at STAR BES-II, CBM@FAIR~\cite{Messchendorp:2025men}, HIAF~\cite{Bai:2026syt}, and future NICA experiments, where large hyperon and deuteron yields and excellent particle identification are available. Since the source size in these collisions is typically of the order of a few femtometers, comparable to the range considered in the present study, the predicted low-momentum correlation functions should be directly accessible experimentally. 

Motivated by these developments, I investigate the femtoscopic correlations of the $d-\Lambda$, $d-\Sigma$, and $d-\Xi$ systems. The effective $d$--$Y$ interactions are constructed by folding hyperon--nucleon potentials over the deuteron wave function, from which the scattering observables and correlation functions are calculated. 

This paper is organized as follows. In
Section~\ref{sec.folding}, I describe the construction of the microscopic folding potentials for the $d-Y$ systems based on the HAL-QCD $Y-N$ interactions. In
Section~\ref{sec.correlations}, I introduce the femtoscopy framework and present the calculated correlation functions for the $d-\Lambda$, $d-\Sigma$, and $d-\Xi$ systems. In
Section~\ref{sec.feeddown}, I investigate the feed-down contributions from excited hyperons to the ground-state $\Lambda$ and their impact on the measured $d-\Lambda$ correlation function. Finally, the conclusion is summarized in Section~\ref{sec.summary}.

\section{Folding potentials for the $dY$ systems}
\label{sec.folding}
The folding potential is a well-established microscopic approach for constructing effective interactions between a composite nucleus and an external hadron~\cite{Satchler:1979ni}. The basic idea is that the interaction between the projectile and the nucleus arises from the coherent sum of the elementary two-body interactions between the projectile and each constituent nucleon. Assuming that the internal structure of the nucleus remains unchanged during the scattering process, the many-body interaction can be reduced to an effective one-body potential by averaging the underlying hadron--nucleon interaction over the nuclear wave function. As a result, the effective nucleus--hadron potential naturally incorporates both the spatial structure of the nucleus and the microscopic dynamics of the elementary interaction.

For a nucleus consisting of $A$ nucleons, the folding potential of $A-Y$ system can generally be written as
\begin{eqnarray}
\mathcal U_{AY}(\mathbf{R})
= \left\langle \Psi_A \left| \sum_{i=1}^{A}
V_{YN}(\mathbf{R}-\mathbf{r}_i) \right|\Psi_A
\right\rangle,
\end{eqnarray}
where $\mathbf{R}$ denotes the relative coordinate between the hyperon and the nuclear center of mass, $\mathbf{r}_i$ is the coordinate of the $i$th nucleon inside the nucleus, $V_{YN}$ is the elementary hyperon--nucleon interaction, and $\Psi_A$ is the nuclear wave function. 

For light nuclear systems, the folding approach is particularly reliable because realistic few-body wave functions can be obtained by solving the corresponding Schr\"odinger equation with high precision. The deuteron, being the lightest bound nucleus, is therefore an ideal candidate for constructing a microscopic hyperon--nucleus interaction. Since its wave function is accurately known from well-known nucleon--nucleon potentials, the resulting deuteron--hyperon interaction can be derived with minimal model dependence. Moreover, because the deuteron is spatially extended and only weakly bound, the folded potential remains directly sensitive to the long-range behavior of the underlying hyperon--nucleon interaction. 
In hypernuclear studies, microscopic folding models have been extensively employed to construct effective $\Lambda$-, $\Sigma$-, and $\Xi$-nucleus interactions from hyperon--nucleon potentials, providing a successful description of hypernuclear binding energies, spectra, and hyperon--nucleus scattering observables~\cite{Cobis:1996ru,Congleton:1992kk,Etminan:2019gds,Etminan:2024ukl,Filikhin:2025ivu,Jinno:2024tjh}.

For the $\Lambda-N$ and $\Sigma-N$ systems, the HAL-QCD
interaction is available in the spin-singlet
$^{1}S_0$ channel and the coupled
$^{3}S_1$--$^{3}D_1$ channel due to the tensor interaction~\cite{Nemura:2017vjc}. 
In the present exploratory calculation, the tensor interaction is neglected and retain only the central components of the HAL-QCD potentials. This
approximation is motivated by the relatively weak tensor force in the hyperon--nucleon interaction and the small $D$-wave probability of the
deuteron. For the $\Xi-N$ interaction, the HAL-QCD only gives the central potential in the
$^{1}S_0$ and $^{3}S_1$ channels,
which does not explicitly include the
$S$--$D$ tensor coupling~\cite{HALQCD:2019wsz}.

The effective deuteron--hyperon ($d-Y$) interactions ($Y=\Lambda,\Sigma,\Xi$) are constructed by folding the
HAL-QCD $YN$ interactions over the deuteron wave function,
\begin{eqnarray}
\mathcal U_{\alpha\beta}^{I,s}(R)
&=& \int d^3\rho \psi_d^\dagger(\boldsymbol{\rho})
\Big[ V_{\alpha\beta}^{I,s}
\!\left( \left| \mathbf R+\frac{\boldsymbol{\rho}}{2} \right| \right) \nonumber\\
&+& V_{\alpha\beta}^{I,s}
\!\left( \left| \mathbf R-\frac{\boldsymbol{\rho}}{2} \right| \right) \Big]
\psi_d(\boldsymbol{\rho}),
\label{eq:folding}
\end{eqnarray}
where $\mathbf R$ denotes the relative coordinate between the hyperon and the
deuteron center of mass, and $\boldsymbol{\rho}$ is the proton--neutron
relative coordinate. $\psi_d$ is deuteron wavefucntion, I take it from the previous study~\cite{Zhao:2025glf}. $\alpha,\beta=\Lambda N,\Sigma N$. The upper index $I$ and $s$ represent the isospin and spin quantum number, respectively.

In this paper, the central $\Lambda-N$, $\Sigma -N$, and $\Xi-N$ interactions are taken from
the HAL-QCD. The $\Lambda-N$ and $\Sigma-N$ potentials are parameterized by a two-Gaussian form,
\begin{equation}
V_{YN}^{I,s}(r) = \sum_{i=1}^{2} \alpha_i \exp\!\left(-\frac{r^2}{\beta_i^2}\right), \qquad (Y=\Lambda,\Sigma),
\end{equation}
whereas the $\Xi-N$ interaction is parameterized as
\begin{eqnarray}
V_{\Xi N}^{I,s}(r)
&=& \sum_{i=1}^{3} \alpha_i
\exp\!\left(-\frac{r^2}{\beta_i^2}\right)
+\lambda_2Y^2(\rho_2,m_\pi,r)\nonumber\\
&+& \lambda_1Y(\rho_1,m_\pi,r),
\end{eqnarray}
with
\begin{eqnarray}
Y(\rho,m,r)=\left(1-e^{-r^2/\rho^2}\right)
\frac{e^{-m_\pi r}}{r}.
\end{eqnarray}
The corresponding parameters for all spin--isospin channels are taken directly from the HAL-QCD parameterizations~\cite{Nemura:2017vjc,HALQCD:2019wsz}.

The deuteron has spin $S_d=1$, whereas all hyperons have $S_Y=\frac12$.
Consequently, the total angular momentum,
\begin{equation}
J=\frac12,\qquad\frac32.
\end{equation}

Since the elementary interaction acts between the hyperon and one nucleon inside the deuteron, the interacting $YN$ pair must first be coupled to definite spin $s=0$ or 1. The transformation from the deuteron coupling scheme $|(NN)S_d,Y;J\rangle$ to the $YN$ coupling scheme $|(YN)s,N;J\rangle$ is accomplished through Wigner $6j$ coefficients,
\begin{eqnarray}
|(NN)S_d,Y;J\rangle=\sum_s A_s |(YN)s,N;J\rangle,
\end{eqnarray}
where
\begin{eqnarray}
A_s = (-1)^{1+J} \sqrt{3(2s+1)}
\left\{
\begin{matrix}
\frac12 & \frac12 & 1\\
\frac12 & J & s
\end{matrix}
\right\}.
\end{eqnarray}
The deuteron is a spin-triplet state $S_d=1$. Since the folded interaction is obtained from the expectation value $\langle \Psi|V_{YN}|\Psi \rangle$, the contribution of each $YN$ spin channel is weighted by the square of the corresponding recoupling amplitude $C_s^{(J)}=|A_s|^2$, leading to
\begin{equation}
C_0^{(1/2)}=\frac34,\qquad
C_1^{(1/2)}=\frac14,
\end{equation}
for $J=1/2$ channel. Whereas for the $J=3/2$ channel only the spin-triplet $YN$
configuration contributes,
\begin{equation}
C_0^{(3/2)}=0,\qquad
C_1^{(3/2)}=1.
\end{equation}
These spin-related coefficients are universal for the $d-\Lambda$, $d-\Sigma$, and $d-\Xi$
systems.

The deuteron is an isosinglet state with $I_d=0$, whereas the $\Lambda$ hyperon is an
isosinglet, $I_\Lambda=0$.
The microscopic $d-\Lambda$ interaction is constructed from the
elementary $\Lambda-N$ interaction, which acts between the $\Lambda$ and one nucleon inside the deuteron, while the second nucleon
acts as a spectator. Since the $\Lambda$ carries zero isospin and the nucleon has
$I_N=1/2$, the interacting $\Lambda-N$ pair can only couple to
\begin{equation}
I_{\Lambda N} = 0\otimes\frac12 = \frac12 .
\end{equation}
Consequently, no isospin recoupling average is required for the $\Lambda N$ interaction.
The effective potentials are therefore
\begin{eqnarray}
U_{d\Lambda}^{J=1/2}
&=&\frac34 \mathcal U_{\Lambda N}^{I=1/2,s=0}
+\frac14 \mathcal U_{\Lambda N}^{I=1/2,s=1},
\nonumber\\
U_{d\Lambda}^{J=3/2}
&=&\mathcal U_{\Lambda N}^{I=1/2,s=1}.
\end{eqnarray}

Different from $\Lambda$, $\Sigma$ carries isospin $I_\Sigma=1$.
Since the interacting $\Sigma-N$ pair consists of particles with
isospins $I_\Sigma=1$ and $I_N=1/2$, it can couple to
\begin{eqnarray}
I_{\Sigma N}=1\otimes\frac12=\frac12\oplus\frac32.
\end{eqnarray}
The transformation from the deuteron coupling scheme
$|(NN)I_d,\Sigma;I\rangle$ to the $\Sigma N$ coupling scheme
$|(\Sigma N)I_{\Sigma N},N;I\rangle$ is described by Wigner $6j$
coefficients, which is similar to the previous  discussed in spin space.
For the deuteron ($I_d=0$) and total isospin $I=1$, the recoupling
coefficients become, $W^{(I)}_{I_{YN}}$,
\begin{eqnarray}
W^{(1)}_{1/2}=\frac13,\qquad
W^{(1)}_{3/2}=\frac23.
\end{eqnarray}
Accordingly, the folded $d-\Sigma$ potential is obtained as the weighted
sum of the the elementary $\Sigma-N$ spin-isospin channels,
\begin{eqnarray}
U_{d\Sigma}^{J=1/2}
&=&\frac34\left(\frac13\mathcal U_{\Sigma N}^{I=1/2,s=0}+\frac23\mathcal U_{\Sigma N}^{I=3/2,s=0}\right)
\nonumber\\
&+&\frac14\left(\frac13\mathcal U_{\Sigma N}^{I=1/2,s=1}+\frac23\mathcal U_{\Sigma N}^{I=3/2,s=1}\right),
\nonumber\\
U_{d\Sigma}^{J=3/2}
&=&\frac13\mathcal U_{\Sigma N}^{I=1/2,s=1}
+\frac23\mathcal U_{\Sigma N}^{I=3/2,s=1}.
\end{eqnarray}

Since the strong interaction conserves total isospin, only channels with identical isospin quantum numbers can couple. Therefore, the $\Lambda N$ channel couples exclusively to the $I=\frac12$ component of the $\Sigma-N$ system, as shown clearly in HAL-QCD simulation.
However, $d-\Lambda$ and $d-\Sigma$ coupling is not allowed, due to the different isospin, $I_{d-\Lambda}\neq I_{d-\Sigma}$. 

$\Xi$ carries isospin $I_\Xi=1/2$. The interacting $\Xi-N$ pair consists of particles with isospins $I_\Xi=1/2$ and $I_N=1/2$, therefore it can couple to
\begin{eqnarray}
I_{\Xi N}=\frac12\otimes\frac12=0\oplus1.
\end{eqnarray}
Following the similar way, I obtain the corresponding isospin recoupling coefficients are
\begin{eqnarray}
W^{(1/2)}_{0}=\frac14,\qquad
W^{(1/2)}_{1}=\frac34.
\end{eqnarray}
The effective potentials are
\begin{eqnarray}
U_{d\Xi}^{J=1/2}
&=&\frac34\left(\frac14\mathcal U_{\Xi N}^{I=0,s=0}+\frac34\mathcal U_{\Xi N}^{I=1,s=0}\right)
\nonumber\\
&+&\frac14\left(\frac14\mathcal U_{\Xi N}^{I=0,s=1}+\frac34\mathcal U_{\Xi N}^{I=1,s=1}\right),
\nonumber\\
U_{d\Xi}^{J=3/2}
&=&\frac14\mathcal U_{\Xi N}^{I=0,s=1}
+\frac34\mathcal U_{\Xi N}^{I=1,s=1}.
\end{eqnarray}
\begin{figure}[!htb]
    \centering
    \includegraphics[width=0.45\textwidth]{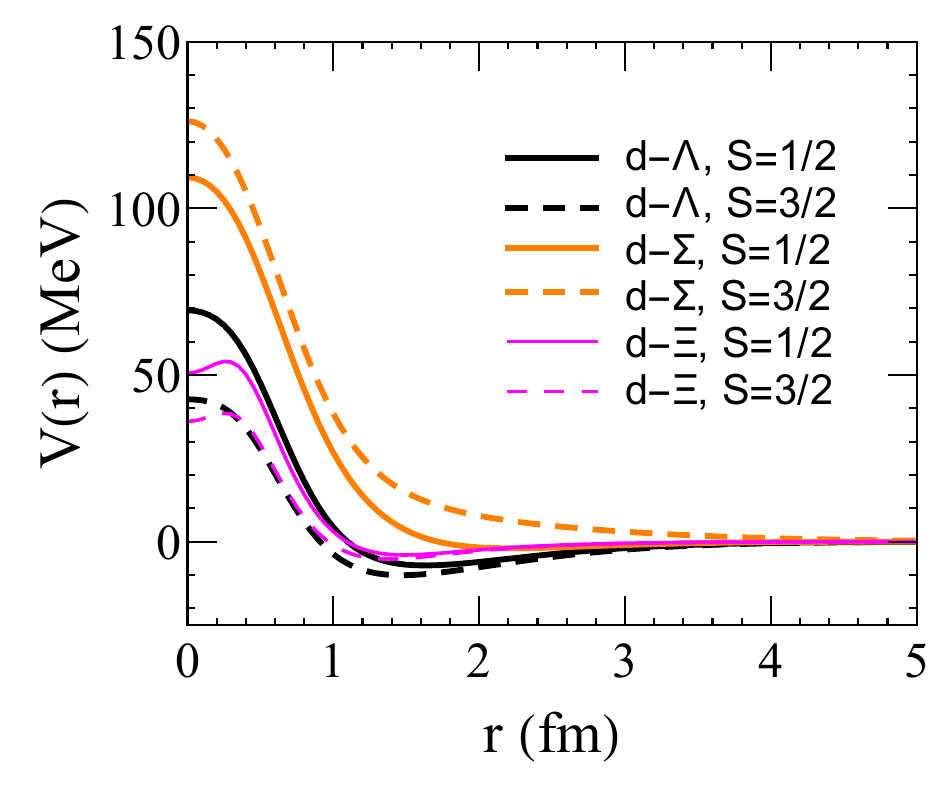} 
    \caption{The strong interactions between $d$ and $Y$ $(Y=\Lambda,\Sigma,\Xi)$. The solid lines represent spin-singlet state, while dashed lines are for spin-triplet state.}
    \label{fig1}
\end{figure}

The folding potential for each $d-Y$ pair and different total spin channels are shown in Fig.~\ref{fig1}. All interactions exhibit a repulsive core at short distances and an attractive pocket at intermediate distances, reflecting the underlying hyperon--nucleon interactions folded with the deuteron wave function. Among them, the $d-\Sigma$ potential is the most repulsive, while the $d-\Xi$ interaction is comparatively weaker. The spin dependence is most pronounced for the $d-\Lambda$ and $d-\Sigma$ systems, where the $J=3/2$ channel becomes more attractive than the $J=1/2$ channel, whereas the difference between the two spin states is much smaller for the $d-\Xi$ system. All potentials approach zero at large distances.

With the $d-Y$ interaction, I solve the two-body Schr\"odinger equation. No bound states are founded for $d-\Lambda$, $d-\Sigma$, and $d-\Xi$ system. 

\section{correlations}
\label{sec.correlations}
Within the femtoscopy formalism, the experimentally observed correlation function can be expressed theoretically as a convolution of the source function $S(\mathbf{r})$ with the relative two-body scattering wave function $\psi_k(\mathbf{r})$~\cite{Koonin:1977fh},
\begin{eqnarray}
C(k)=\int S({\bm r})|\Psi_k({\bm r})|^2d{\bm r},
\label{eq.correlation}
\end{eqnarray}
where $k=|{\bf p}_1^*-{\bf p}_2^*|/2$ is the relative momentum in the center-of-mass frame of the pair, and ${\bf r}$ is the relative distance between the two particles. $\Psi_k({\bf r})$ is the relative wavefucntion of two-body and can be expressed as 
\begin{eqnarray}
\Psi_k(\bm{r})=e^{i\bm{k}\cdot\bm{r}} +\sum_{l=0}^{l_{\rm max}}(2l+1)i^l\left[\frac{u_{k,l}(r)}{kr}-j_l(kr)\right]P_l,
\end{eqnarray}
where $P_l$ and $j_l$ denote the Legendre polynomials and the spherical Bessel function of the first kind of order $l$, respectively. $u_{k,l}(r)$ is reduced radial scattering wave function for the $l$-th partial wave, which can be calculated by solving the Schr\"odinger equation with the aforementioned potential. The radial Schr\"odinger equation for the scattering states is,
\begin{eqnarray}
\frac{d^2 u_{k,l}(r)}{dr^2} = \left( 2\mu_{dY} U_{dY}^{J}(r) + \frac{l(l+1)}{r^2} - k^2 \right) u_{k,l}(r),
\end{eqnarray}
where $\mu_{dY} = (m_dm_Y) / (m_d+m_Y)$ is the reduced mass. For low-energy scattering, the $S-$wave channel dominates. However, as the energy increases, contributions from larger angular momenta become significant. For low-energy scattering, the series converges relatively rapidly. I checked the scattering wavefunction I obtained is identical to the wavefunction from the CATS with the same potential~\cite{Mihaylov:2018rva,Fabbietti:2020bfg}. 

\begin{table}[!bt]
\renewcommand\arraystretch{1.8}
\caption{Comparison the scattering length $f_0$ and interaction range $d_0$ of $d-\Lambda$ system from my calculation, experiment, and other models. The unit is fm.}
\label{tab1}
\setlength{\tabcolsep}{0.5mm}
\begin{tabular}{c|c|c|c|c}
	\toprule[1pt]\toprule[1pt] 
	$d-\Lambda$& $f_0^{J={1/2}}$ & $d_0^{J={1/2}}$  &$f_0^{J={3/2}}$& $d_0^{J={3/2}}$  \\
    \toprule[1pt] 
    This study &-3.5 & 5.5 & -13.6  & 3.8
	\\  
    \toprule[1pt] 
    Exp.~\cite{STAR:2025jwe} & $-26.1\pm 5.6$ & 8 & $18.7\pm 2.8$ &  $6.5\pm 1.8$
	\\  
    \toprule[1pt] 
    Cobis~\cite{Cobis:1996ru} & -16.8 & 3.2 & - & -
	\\
    \toprule[1pt]
    Hammer~\cite{Hammer:2001ng}& -16.8 & 2.3 & -  & -
	\\  
	\bottomrule[1pt]\bottomrule[1pt]
\end{tabular}
\end{table}

The $S$-wave scattering phase shift is obtained by matching the
numerical radial wave function to its asymptotic form,
\begin{equation}
u_{k,0}(r)
\xrightarrow[r\rightarrow\infty]{} \sin\left(kr+\delta_0\right),
\end{equation}
where $\delta_0(k)$ denotes the $S$-wave phase shift. The scattering length $f_0$ and effective range
$d_0$ are then extracted from the effective-range expansion,
\begin{equation}
k\cot\delta_0(k)=-\frac{1}{f_0}+\frac12 d_0k^2
+\mathcal{O}(k^4).
\end{equation}
The scattering length $f_0$ and interaction range $d_0$ of $d-\Lambda$ system is calculated and compared with other results in Table~\ref{tab1}.
Although the folded $d-\Lambda$ potentials are attractive, neither the
$J=1/2$ nor the $J=3/2$ channel supports a two-body bound state in the
present calculation, which is consistent with the bound state Schr\"odinger equation calculation.

\begin{figure}[!htb]
    \centering
    \includegraphics[width=0.45\textwidth]{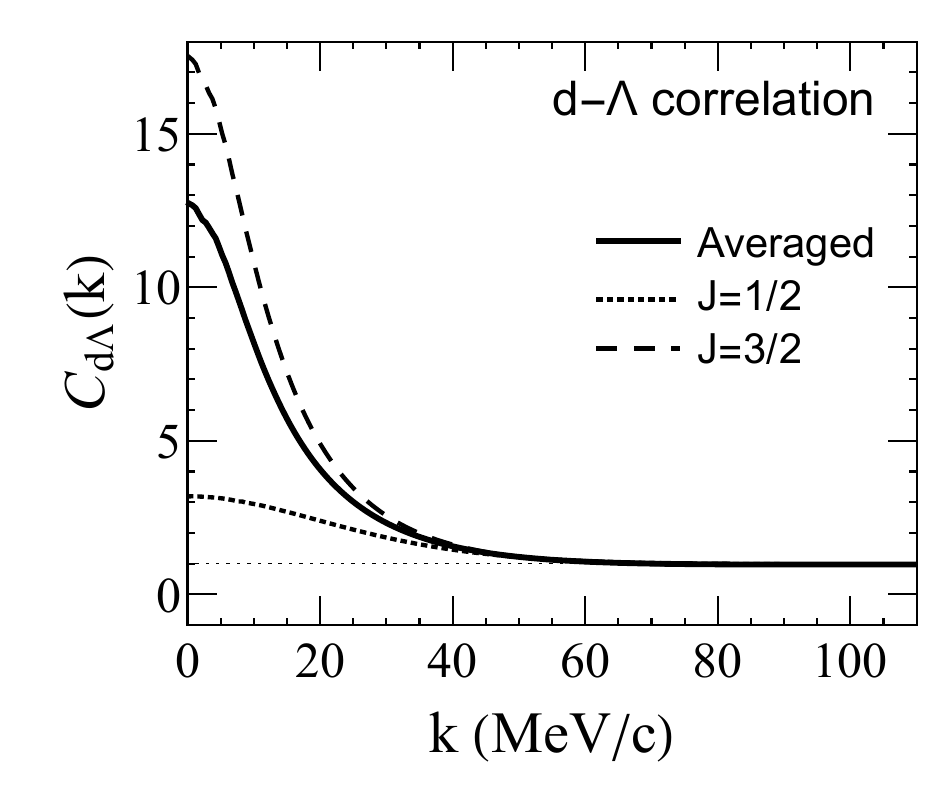} 
    \includegraphics[width=0.45\textwidth]{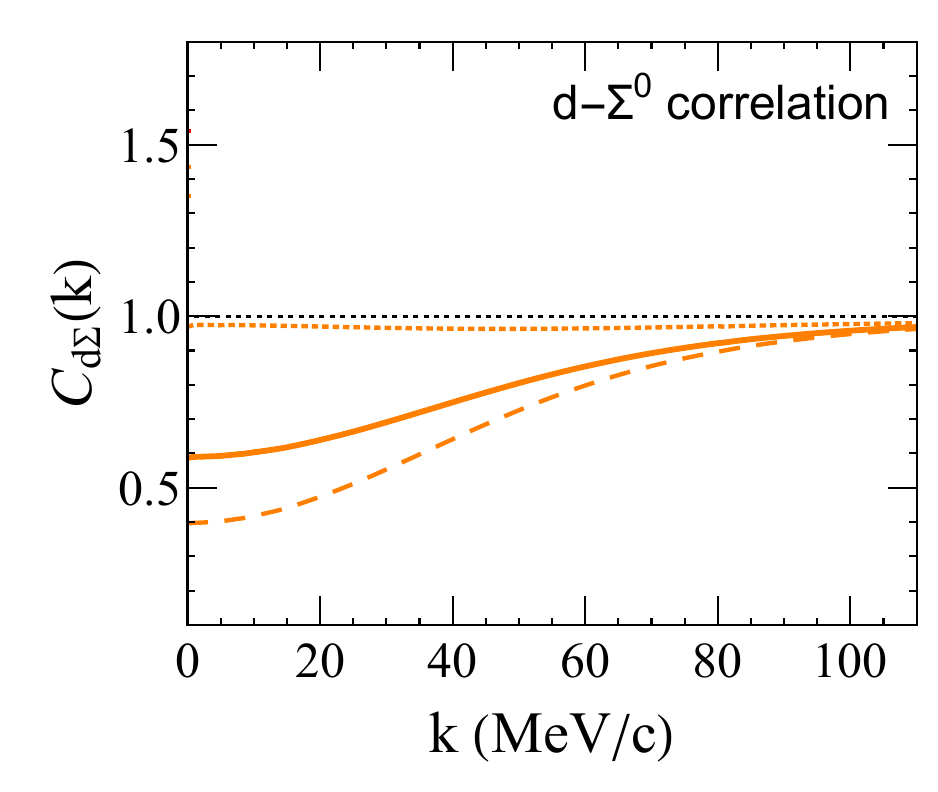} 
    \caption{The correlations function of $d-\Lambda$ (upper panel), $d-\Sigma^0$ (lower panel). The source parameter $r_0=2.3~\rm fm$.}
    \label{fig2}
\end{figure}

Now, let us come to calculate the correlations with a assumption of an isotropic Gaussian emission source, 
\begin{eqnarray}
S(r) = {1\over (4\pi r_{0}^{2})^{3/2}}\exp\left(-{r^{2}\over 4r_{0}^{2}}\right),
\end{eqnarray}
where $r_0$ characterizes the source size. 
For each $dY$ pair, the measured correlation function is the spin-averaged value,
\begin{eqnarray}
C_{dY}(k)={1\over3 }C_{S={1/2}}(k)+{2\over3}C_{S={3/2}}(k),
\end{eqnarray}
where $C_{S={1/2}}(k)$ and $C_{S={3/2}}(k)$ are correlation functions for spin 1/2 and 3/2 channels, respectively. The correlations of each spin component and spin-averaged value are shown in Fig.~\ref{fig2} for $d-\Lambda$ and $d-\Sigma$. Both spin channels exhibit a pronounced
enhancement at small relative momentum due to the attractive
$d-\Lambda$ final-state interaction. The enhancement is considerably
stronger in the $J=3/2$ channel, where the correlation function reaches
its maximum at $k\rightarrow0$. This behavior originates from the much
larger magnitude of the scattering length,
$|f_0^{J=3/2}|=13.6~\mathrm{fm}$, compared with
$|f_0^{J=1/2}|=3.5~\mathrm{fm}$. The large scattering length indicates
that the $J=3/2$ $d-\Lambda$ system is close to a near-threshold pole,
which significantly enhances the scattering wave function at low
relative momentum and consequently produces a strong femtoscopic
correlation. The spin-averaged correlation lies between the two
individual spin channels. For $d-\Sigma$ system, 
due to the stronger repulsive potential and almost no attractive, the correlation is smaller than 1 as shown in the lower panel of Fig.~\ref{fig2}. 

\begin{figure}[!htb]
    \centering
    \includegraphics[width=0.45\textwidth]{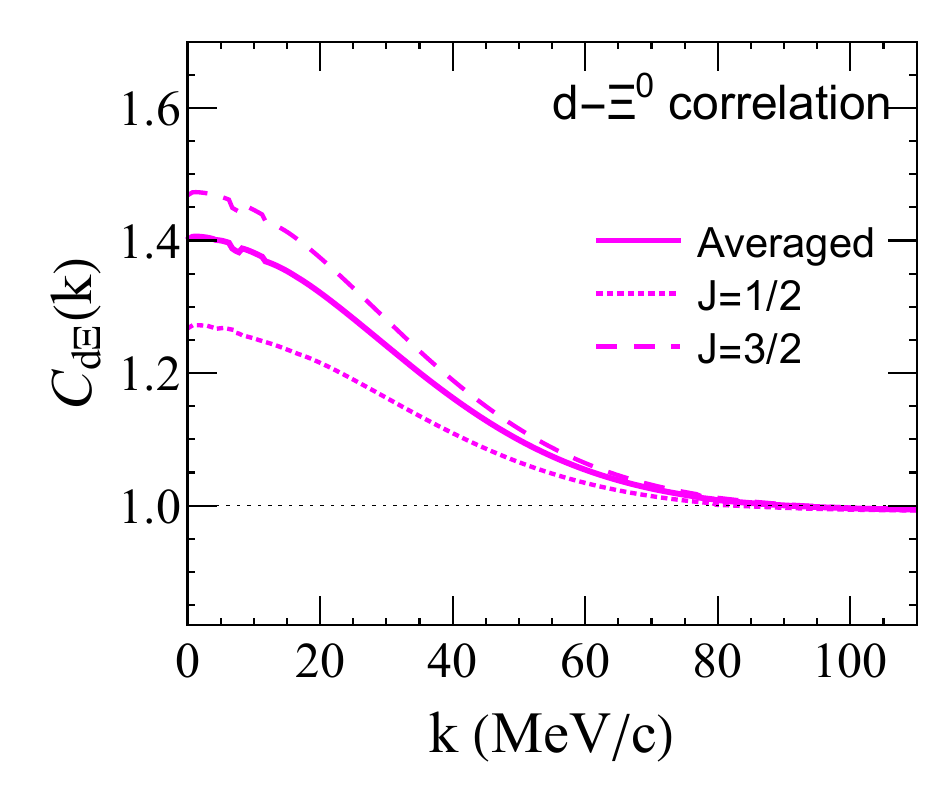} 
    \includegraphics[width=0.45\textwidth]{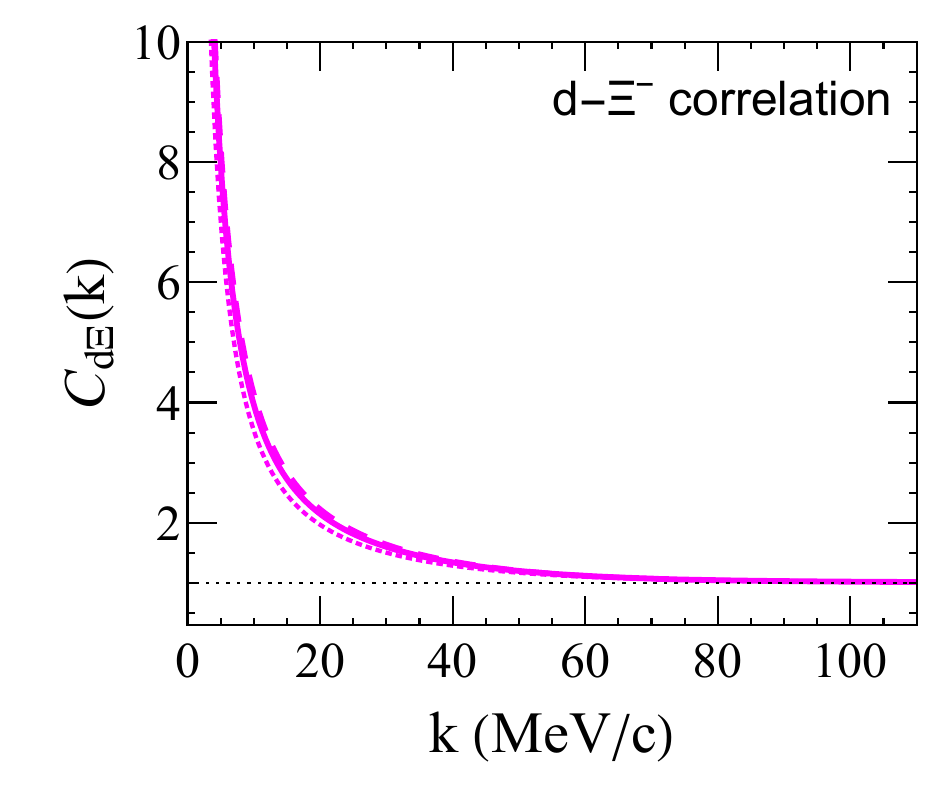} 
    \caption{The correlations function of $d-\Xi^0$ (upper panel) and $d-\Xi^-$ (lower panel). The source parameter $r_0=2.3~\rm fm$.}
    \label{fig2-2}
\end{figure}

Fig.~\ref{fig2-2} presents the calculated $d-\Xi^{0}$ and
$d-\Xi^{-}$ correlation functions for the $J=1/2$ and $J=3/2$ channels
together with the corresponding spin-averaged results. For the neutral
$d-\Xi^{0}$ system, the correlation function exhibits only a moderate
enhancement at small relative momentum. This behavior reflects the
relatively weak attractive strong interaction. Consequently, the difference between
the two spin channels remains modest, with the $J=3/2$ channel showing
a slightly stronger enhancement owing to its somewhat larger attraction.

In contrast, the $d-\Xi^{-}$ correlation function displays a pronounced
enhancement at very small relative momentum. Since the strong
interactions in the $d-\Xi^{0}$ and $d-\Xi^{-}$ systems are identical, this additional enhancement originates primarily from the
long-range attractive Coulomb interaction between the positively
charged deuteron and the negatively charged $\Xi^{-}$ hyperon. The
Coulomb attraction significantly increases the probability of finding
the two particles at small relative momentum, resulting in a sharp rise
of the correlation function near $k=0$. At larger relative momenta, the
Coulomb effect becomes progressively weaker, and the correlation
function gradually approaches unity.

\section{Feed-down contributions}
\label{sec.feeddown}
Besides primordial produced $d\Lambda$ pairs, the experimentally measured
$d-\Lambda$ correlation function also receives contributions from the
decays of heavier hyperons, including mostly
$\Sigma^{0}\rightarrow\Lambda\gamma$,
$\Sigma(1385)\rightarrow\Lambda\pi$,
and
$\Xi\rightarrow\Lambda\pi$~\cite{ParticleDataGroup:2022pth}.
Since the lifetimes of these parent hyperons are much longer than the timescale of the strong final-state interaction, the parent $dY$
($Y=\Sigma^{0},\,\Sigma(1385),\,\Xi$) pair first develops its femtoscopic correlation before the hyperon decays.
The decay subsequently changes the relative momentum of the daughter $d\Lambda$ pair, thereby redistributing the parent correlation over the
measured $d\Lambda$ momentum.

The feed-down contribution from $\Sigma^{0}$ can be treated explicitly,
since the $d-\Sigma^{0}$ interaction is calculated in the present work and the subsequent electromagnetic decay $\Sigma^{0}\rightarrow\Lambda\gamma$ is well understood. The corresponding response matrix can be obtained by Monte Carlo simulation
of the decay kinematics.

The treatment of the $\Sigma(1385)$ feed-down is less straightforward.
At present, no reliable $\Sigma(1385)-N$ interaction is available from
either lattice QCD or phenomenological models, preventing the
construction of a realistic $d-\Sigma(1385)$ potential.
Therefore, the baseline calculation neglects the corresponding
final-state interaction and assumes $C_{d\Sigma(1385)}(k)=1$. 
To estimate the uncertainty associated with this approximation, I also
consider an alternative scenario in which the unknown
$d-\Sigma(1385)$ interaction is approximated by the
$J=3/2$ $d\Sigma$ correlation,
$C_{d\Sigma(1385)}(k) \simeq
C_{d\Sigma}^{J=3/2}(k)$,
motivated by the fact that the $\Sigma(1385)$ is a spin-$3/2$
resonance. The difference between these two prescriptions is taken as a
systematic uncertainty associated with the unknown
$\Sigma(1385)-N$ interaction.

Finally, the contribution from $\Xi$ feed-down depends on the
experimental reconstruction procedure. Since the weak decay
$\Xi\rightarrow\Lambda\pi$ occurs far outside the femtoscopic region,
its contribution should be included only if secondary $\Lambda$
hyperons from $\Xi$ decays are retained in the experimental sample.
Otherwise, it is naturally removed by the selections used
to reject weak-decay backgrounds.

Now, let us come to construct the decay kinematics by means of a Monte Carlo simulation.
For a parent hyperon $Y$ with mass $m_Y$, a parent
$dY$ pair is first generated in its pair rest frame with a fixed relative
momentum $k_i$,
\begin{equation}
p_Y^\mu=(E_Y,0,0,+k_i),
\qquad
p_d^\mu=(E_d,0,0,-k_i),
\end{equation}
where
\begin{equation}
E_Y=\sqrt{m_Y^2+k_i^2},
\qquad
E_d=\sqrt{m_d^2+k_i^2}.
\end{equation}

The hyperon decay is then simulated in the hyperon rest frame.
For the two-body decays
$\Sigma^{0}\rightarrow\Lambda\gamma$,
$\Sigma(1385)\rightarrow\Lambda\pi$ (here include $\Sigma(1385)^0, \Sigma(1385)^\pm$),
and
$\Xi\rightarrow\Lambda\pi$ (here include $\Xi^0, \Xi^-$),
the daughter $\Lambda$ is emitted isotropically with a fixed momentum
\begin{equation}
q=\frac{\lambda^{1/2}(m_Y^2,m_\Lambda^2,m_X^2)}
{2m_Y},
\end{equation}
where $m_X$ denotes the mass of the emitted photon or pion and $\lambda(x,y,z)=x^2+y^2+z^2-2xy-2xz-2yz$
is the K\"all\'en function.
The emission direction is sampled uniformly over the solid angle,
\begin{equation}
\cos\theta\in[-1,1],
\qquad
\phi\in[0,2\pi),
\end{equation}
leading to the daughter four-momentum
\begin{equation}
p_\Lambda^\mu
=
(E_\Lambda,
q\sin\theta\cos\phi,
q\sin\theta\sin\phi,
q\cos\theta),
\end{equation}
with $E_\Lambda=\sqrt{m_\Lambda^2+q^2}$.

The daughter $\Lambda$ is subsequently boosted back into the parent $dY$ pair rest frame according to the parent hyperon velocity. The relative momentum of the reconstructed $d\Lambda$ pair, denoted by $k_f$, is then calculated in the $d\Lambda$ pair rest frame. Repeating this procedure for a large number of Monte Carlo events yields the response matrix $T(k_f,k_i)$, which represents the probability that a parent pair produced with relative momentum $k_i$ is reconstructed as a daughter pair with relative momentum $k_f$. The response matrix is normalized column by column,
\begin{equation}
\sum_{k_f}T(k_f,k_i)=1,
\end{equation}
ensuring probability conservation for each parent momentum.

The feed-down contribution to the observed $d-\Lambda$ correlation is
obtained by folding the parent correlation function with the response
matrix,
\begin{equation}
C_Y^{\rm fd}(k_f)
=
\frac{
\sum_i
T(k_f,k_i)
A(k_i)
C_{dY}(k_i)
}{
\sum_i
T(k_f,k_i)
A(k_i)
},
\end{equation}
where $C_{dY}(k_i)$ denotes the parent correlation function and
$A(k_i)$ is the parent momentum distribution.
In the present work, a uniform parent momentum distribution,
$A(k_i)=1$, is adopted.
Finally, the experimentally observed $d-\Lambda$ correlation is obtained
as the weighted sum of the primordial and feed-down contributions,
\begin{equation}
C_{d\Lambda}^{\rm obs}(k)=
f_{\rm pri} C_{d\Lambda}(k) + \sum_Y f_Y C_Y^{\rm fd}(k),
\end{equation}
where $f_{\rm pri}$ and $f_Y$ denote the fractions of primordial produced
$\Lambda$ hyperons and those originating from hyperon decays, respectively. The total fraction is unit, $f_{\rm pri}+\sum f_Y=1$. In this study, I take the fraction from the thermal model for Au+Au collisions at $\sqrt{s_{\rm NN}}=3~\rm GeV$, which gives, $f_{\rm dir}\approx 0.523$, $f_\Sigma^0 \approx 0.242$, $f_{\Sigma^{*0}}+f_{\Sigma^{*\pm}}\approx 0.195$, and $f_{\Xi^0} + f_{\Xi^-} \approx 0.04$~\cite{Vovchenko:2015idt,Vovchenko:2019pjl}. For other colliding energy and system, these ratios maybe different.

\begin{figure}[!htb]
    \centering
    \includegraphics[width=0.45\textwidth]{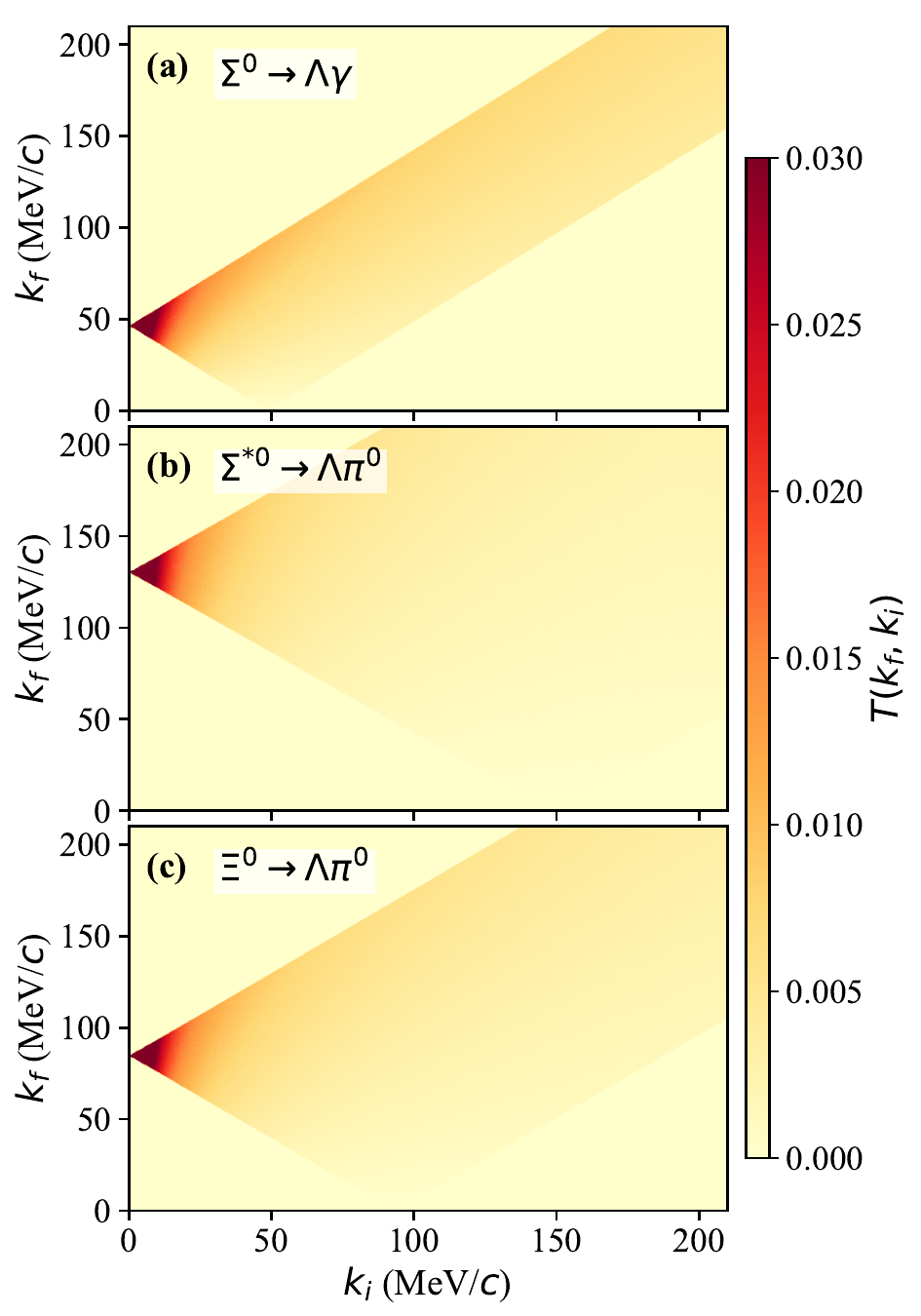} 
    \caption{the response matrix $T(k_f,k_i)$ for feed-down contributions from $\Sigma^0$ (a), $\Sigma(1385)^0$($\Sigma^{*0}$) (b), and $\Xi^0$ (c) to $\Lambda$.}
    \label{fig3}
\end{figure}

The response matrices
$T(k_f,k_i)$ for the feed-down contributions from
(a) $\Sigma^{0}\rightarrow\Lambda\gamma$,
(b) $\Sigma^{*0}\rightarrow\Lambda\pi^{0}$,
and (c) $\Xi^{0}\rightarrow\Lambda\pi^{0}$ are presented in Fig.~\ref{fig3}. 
Here, $k_i$ and $k_f$ denote the relative momenta of the parent
$dY$ pair and the daughter $d\Lambda$ pair, respectively.
The three decay channels exhibit distinct kinematic response patterns,
which are governed by their decay momenta in the parent rest frame.
For the electromagnetic decay
$\Sigma^{0}\rightarrow\Lambda\gamma$,
the recoil momentum is relatively small
($q\simeq74~\mathrm{MeV}/c$),
leading to a narrow response concentrated close to the diagonal.
In contrast, the strong decay
$\Sigma(1385)^{0}\rightarrow\Lambda\pi^{0}$
has a much larger decay momentum
($q\simeq208~\mathrm{MeV}/c$),
which results in a significantly broader redistribution of the parent relative momentum into the observed $d\Lambda$ momentum. The response matrix for $\Xi^{0}\rightarrow\Lambda\pi^{0}$
shows an intermediate behavior, reflecting its decay momentum
($q\simeq136~\mathrm{MeV}/c$).
Consequently, the feed-down smearing is weakest for
$\Sigma^{0}$, strongest for $\Sigma(1385)^{0}$, and intermediate for $\Xi^{0}$. Another characteristic feature is the finite intercept of the response
matrices at $k_i=0$. Even when the parent $dY$ pair is produced with vanishing relative momentum, the decay recoil imparts a finite momentum to the daughter $\Lambda$, shifting the reconstructed $d\Lambda$ pair.

\begin{figure}[!htb]
    \centering
    \includegraphics[width=0.45\textwidth]{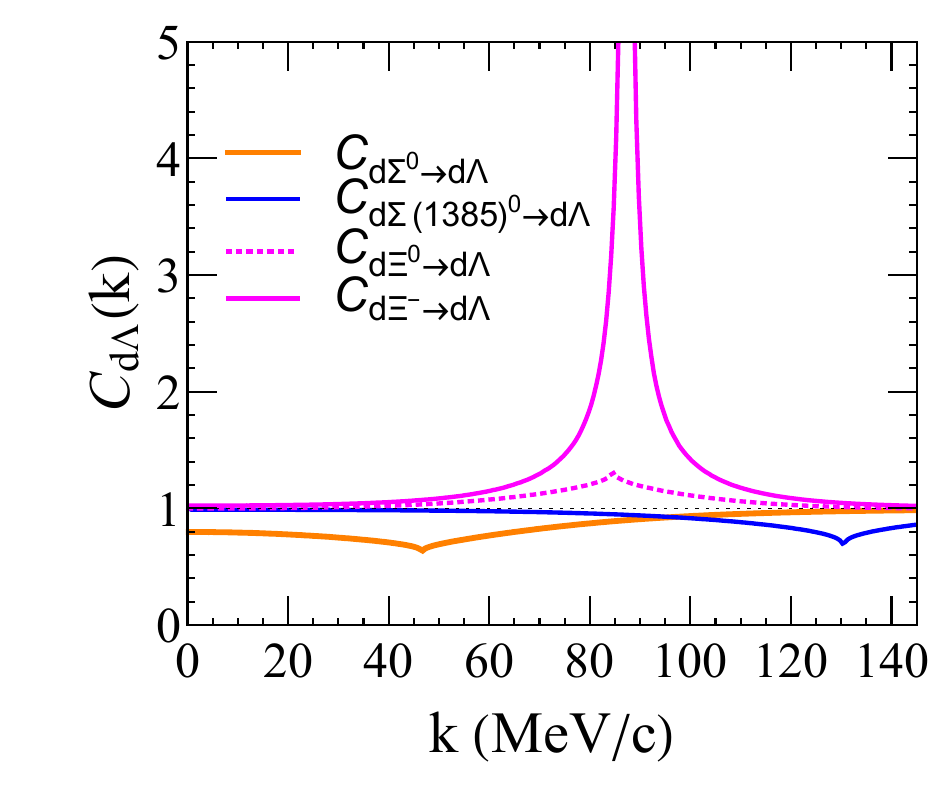} 
    \caption{The feed-down contribution of $d-\Sigma^0$, $d-\Sigma(1385)^0$, $d-\Xi^0$, and $d-\Xi^-$ to $d-\Lambda$ correlations.}
    \label{fig4-0}
\end{figure}
Fig.~\ref{fig4-0} presents the feed-down contributions to the
measured $d-\Lambda$ correlation from the decays
$\Sigma^{0}\rightarrow\Lambda\gamma$,
$\Sigma(1385)^{0}\rightarrow\Lambda\pi^0$,
$\Xi^{0}\rightarrow\Lambda\pi$, and
$\Xi^{-}\rightarrow\Lambda\pi$.
The decay kinematics redistribute the parent $dY$ correlation over the daughter $d\Lambda$ relative momentum, leading to characteristic modifications for different parent hyperons.
The feed-down contribution from $\Sigma^{0}$ is slightly below unity over the entire momentum range. This behavior originates from the moderate $d\Sigma^{0}$ interaction together with the photon recoil in the electromagnetic decay, which smears the parent correlation and
reduces its strength in the observed $d\Lambda$ channel.
For the heavier excited state $\Sigma(1385)$, assuming $C_{d\Sigma(1385)}(k)=1$ leads to a feed-down contribution that remains unity. Alternatively, if the correlation function is approximated by $C_{d\Sigma(1385)}(k)\simeq C_{d\Sigma}^{J=3/2}(k)$, the resulting feed-down correlation is found to be slightly below unity, as shown by the blue curve in Fig.~\ref{fig4-0}.

The $\Xi^{0}$ feed-down remains close to unity, showing only a weak enhancement around $k\simeq85~\mathrm{MeV}/c$. This reflects the relatively weak strong interaction in the $d-\Xi^{0}$ system. In contrast, the $d\Xi^{-}\rightarrow
d\Lambda$ contribution exhibits a pronounced peak near
$k\simeq90~\mathrm{MeV}/c$. This structure originates from the strong Coulomb enhancement in the parent $d-\Xi^{-}$ correlation at very small relative momentum. After the weak decay $\Xi^{-}\rightarrow\Lambda\pi$, the Coulomb-induced low-momentum enhancement is shifted to finite daughter momentum by the decay kinematics, giving rise to the narrow peak observed in the $d-\Lambda$ correlation. 

\begin{figure}[!htb]
    \centering
    \includegraphics[width=0.45\textwidth]{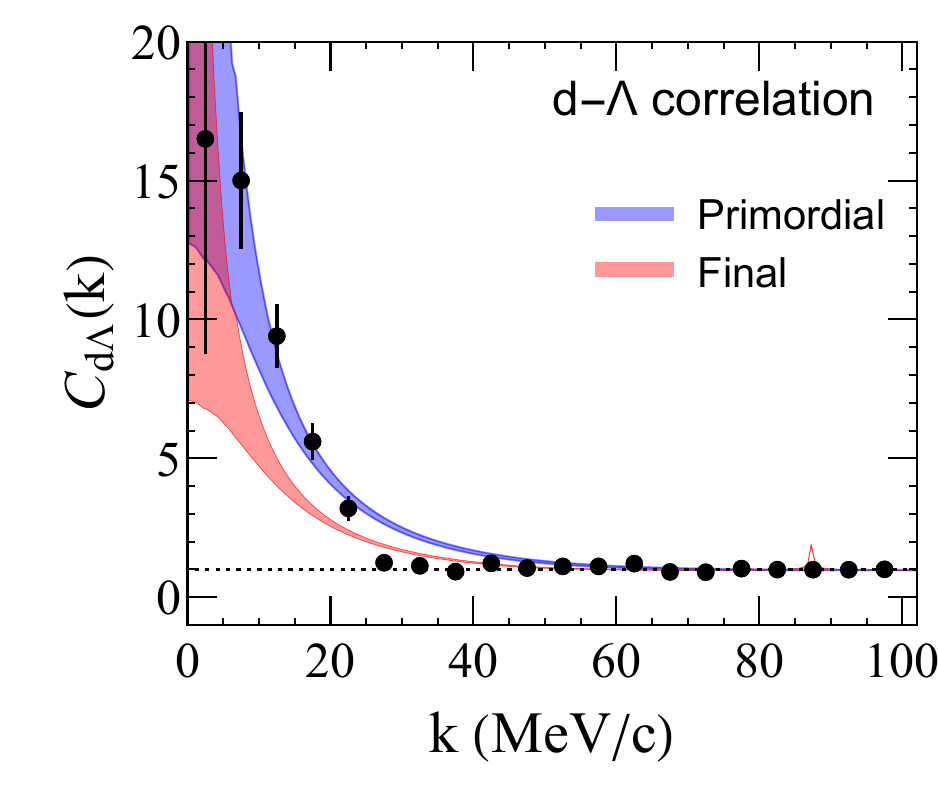} 
    \caption{The correlations of primordial $d-\Lambda$ (blue band) and with feed-down contributions (red band). The data is from STAR measurement in 10-20\% Au+Au collisions at $\sqrt{s_{\rm NN}}=3~\rm GeV$~\cite{STAR:2025jwe}. The shaded bands represent the uncertainty associated with the source size, which varies from $r_0=2.0$ to $2.3$~fm}
    \label{fig4}
\end{figure}

Finally, I compare the primordial and the final
$d-\Lambda$ correlation after including the feed-down contributions from heavier strange baryons, as shown in Fig.~\ref{fig4}. The shaded bands represent
the uncertainty associated with the source size, obtained by varying the Gaussian source radius from $r_0=2.0$ to $2.3$~fm~\cite{STAR:2025jwe}.

The primordial correlation exhibits a strong enhancement at small relative momentum, reflecting the attractive $d\Lambda$
final-state interaction, particularly the large contribution from the $J=3/2$ channel. After incorporating the feed-down contributions, the low-momentum enhancement is noticeably reduced. This suppression
originates primarily from the momentum smearing induced by the decays of the parent hyperons, especially the
$\Sigma^{0}\rightarrow\Lambda\gamma$ decay, which redistributes the parent correlation over a broader momentum region. Although the $\Xi^{-}$ feed-down generates a localized enhancement around
$k\simeq90~\mathrm{MeV}/c$, its contribution is confined to a narrow momentum interval and has only a minor influence on the overall $d-\Lambda$ correlation due to the small fraction $f_{\Xi}$.

The uncertainty associated with the source size is most pronounced at very small relative momenta, where the correlation is most sensitive to the spatial extent of the emission source. As the relative momentum
increases, both the influence of the final-state interaction and the source-size dependence rapidly decrease, and the correlation functions approach unity above
$k\approx40$--$50~\mathrm{MeV}/c$. The final correlation remains substantially enhanced at low relative momentum, indicating that the
$d-\Lambda$ interaction can still be extracted experimentally after properly accounting for feed-down effects.

\section{Summary}
\label{sec.summary}
In this paper, I constructed microscopic $d–Y$ folding potentials from HAL-QCD hyperon–nucleon interactions by explicitly incorporating spin and isospin recoupling. The resulting interactions were used to calculate low-energy scattering observables and femtoscopic correlation functions for the $d-\Lambda$, $d-\Sigma$, and $d-\Xi$ systems. Although no two-body bound states were found, the $d-\Lambda$ system exhibits a pronounced low-momentum enhancement arising from its large scattering length and the proximity of a near-threshold pole. In contrast, the $d-\Sigma$ correlation is suppressed by the predominantly repulsive interaction, while the $d-\Xi^0$ and $d-\Xi^-$ systems are characterized by moderate strong-interaction effects and a substantial Coulomb enhancement, respectively. I further developed a complete feed-down framework including $\Sigma^0$, $\Sigma(1385)$, and $\Xi$ decays through Monte Carlo response matrices. Feed-down modifies the observable $d-\Lambda$ correlation, but the interaction signal remains clearly visible after reconstruction. 

Although the present calculation reproduces the STAR $d-\Lambda$ correlation reasonably well, it slightly underpredicts the extracted scattering length and the low-momentum enhancement. This modest discrepancy may originate from several sources. First, the folded $d-\Lambda$ potential is constructed directly from the current HAL-QCD $\Lambda-N$ interaction without introducing any additional free parameters. Consequently, the strength of the effective $d-\Lambda$ interaction is entirely determined by the underlying $\Lambda-N$ potential. However, the HAL-QCD interaction itself still carries considerable uncertainties, while only its central values are adopted in the present work. If the physical low-energy $\Lambda-N$ interaction is slightly more attractive than the current central HAL-QCD result, the resulting folded $d-\Lambda$ interaction would naturally produce a larger scattering length and a stronger low-momentum femtoscopic correlation. Second, the present calculation is based on the folding approximation, which reduces the genuine three-body $\Lambda pn$ problem to an effective two-body $d-\Lambda$ interaction. Although this approach provides a transparent microscopic description, it neglects explicit three-body dynamics that may become increasingly important for the weakly bound hypertriton~\cite{Viviani:2023kxw,Mrowczynski:2025qys}. Finally, the extraction of the low-energy $d-\Lambda$ scattering parameters from the measured correlation function is itself subject to systematic uncertainties associated with feed-down contributions from excited strange baryons, whose production fractions and residual correlations are not yet precisely constrained. Future measurements with improved statistics and better control of feed-down effects, together with updated lattice-QCD $\Lambda-N$ interactions and fully microscopic three-body calculations, will provide a more stringent benchmark for understanding the low-energy $d-\Lambda$ interaction and the structure of the hypertriton.

\section{Acknowledgements}
The author thanks Yingjie Zhou, Elena Bratkovskaya, and Joerg Aichelin for useful discussions. 
This work acknowledges the support by the Deutsche Forschungsgemeinschaft (DFG) through the grant CRC-TR 211 "Strong-interaction matter under extreme conditions" (Project number 315477589 - TRR 211). 

\bibliographystyle{apsrev4-2}
\bibliography{refs}

\end{document}